\documentclass[conference]{IEEEtran}
\usepackage[space]{cite}
\usepackage{algorithmic}
\usepackage{graphicx}
\usepackage{textcomp}
\usepackage{graphicx}
\usepackage{color}
\usepackage{amsmath,amssymb,amsopn,amstext,amsfonts}
\usepackage{mathtools}
\usepackage{indentfirst}
\usepackage{subfigure}
\usepackage{multirow}
\usepackage{makecell}
\usepackage{comment}
\usepackage{etoolbox}
\usepackage{url}
\usepackage{caption}
\graphicspath{ {./figures/} }

\DeclareGraphicsExtensions{.pdf,.png,.jpg,.eps,.PNG}

\begin{document}

\title{Machine Learning Based Routing Congestion Prediction in FPGA High-Level Synthesis\vspace{-0.25cm}
}
\author{\IEEEauthorblockN{Jieru Zhao$^\ast$, Tingyuan Liang$^\ast$, Sharad Sinha$^\dagger$ and Wei Zhang$^\ast$}
\IEEEauthorblockA{{$^\ast$Department of ECE, Hong Kong University of Science and Technology,} \{jzhaoao,tliang,wei.zhang\}@ust.hk
\\{$^\dagger$Department of CSE, Indian Institute of Technology Goa, sharad\_sinha@ieee.org}}}

\makeatletter
\patchcmd{\@maketitle}
  {\addvspace{0.5\baselineskip}\egroup}
  {\addvspace{-0.65\baselineskip}\egroup}
\makeatother

\maketitle

\begin{abstract}
High-level synthesis (HLS) shortens the development time of hardware designs and enables faster design space exploration at a higher abstraction level. Optimization of complex applications in HLS is challenging due to the effects of implementation issues such as routing congestion. Routing congestion estimation is absent or inaccurate in existing HLS design methods and tools. Early and accurate congestion estimation is of great benefit to guide the optimization in HLS and improve the efficiency of implementation. However, routability, a serious concern in FPGA designs, has been difficult to evaluate in HLS without analyzing post-implementation details after Place and Route. To this end, we propose a novel method to predict routing congestion in HLS using machine learning and map the expected congested regions in the design to the relevant high-level source code. This is greatly beneficial in early identification of routability oriented bottlenecks in the high-level source code without running time-consuming register-transfer level (RTL) implementation flow. Experiments demonstrate that our approach accurately estimates vertical and horizontal routing congestion with errors of 6.71\% and 10.05\% respectively. By presenting \textit{Face Detection} application as a case study, we show that by discovering the bottlenecks in high-level source code, routing congestion can be easily and quickly resolved compared to the efforts involved in RTL level implementation and design feedback.
\end{abstract}
\vspace{-0.15cm}
\section{Introduction}
\vspace{-0.15cm}
Field-programmable gate arrays (FPGAs) can be reprogrammed to implement various tasks with low power and massive parallelism. Due to increasing demands of FPGA applications, high-level synthesis (HLS) becomes a superior alternative to register-transfer level (RTL) methods for its higher productivity. By generating RTL from behavioral descriptions, HLS tools provide synthesis directives and enable fast design space exploration to meet design requirements\cite{zhao2017comba}. However, it is much harder to grasp the hardware implementation details when designing at a higher level of abstraction. Although HLS tools provide estimations of several post-implementation metrics, the estimated values deviate significantly from the actual values because the subsequent optimization in the downstream implementation steps is not modeled\cite{dai2018fast,zheng2014fast}. This inaccuracy affects the correct evaluation of the design quality, misleads designers during the performance optimization process and incurs implementation issues that degrade the final performance. One of the crucial issues is routing congestion.\\
\indent Routing is an important problem in FPGA designs since it contributes a lot to delay and resource utilization. In a congested design, wires have to be detoured for connections, generating longer delays and occupying more routing resources\cite{belghadr2014metro}. Routing congestion degrades the design performance and even leads to implementation failures. Therefore, it is vital that routing congestion could be identified and resolved early in the design cycle. Detection of routing congestion at the HLS source-code level could reduce the iterations of the design cycle and help designers choose a proper optimization scheme. \\
\indent In physical design, many works\cite{chen2017ripplefpga,maaroufmachine,li2018utplacef} predict routing congestion to guide FPGA placement. With informative physical metrics, routing congestion can be predicted with high accuracy. However, as the abstraction level increases, the difficulty of congestion estimation is exacerbated due to the lack of physical information.
Several HLS-based methods \cite{wu2008congestion,dougherty2000unifying,wang2003reallocation,cong2012towards} are proposed to improve HLS scheduling or allocation algorithms to generate layout-friendly RTL models.  \cite{wu2008congestion,dougherty2000unifying,wang2003reallocation} incorporate floorplanning to HLS and \cite{cong2012towards} summarizes multiple RTL metrics to evaluate the quality of HLS-generated models. All of these methods aim at improving HLS algorithms, which is different from our problem that how to detect and eliminate congestion issues in the source code. \cite{tatsuoka2015physically} identifies congestion in HLS by integrating a physically aware logic synthesis tool from Cadence. The required physical information is obtained from the generated congestion reports and no estimation is needed. Since we focus on FPGA-based designs and utilize FPGA commercial design tools like Vivado, related congestion metrics can only be obtained after Place and Route (PAR).\\
\indent To this end, we propose a machine-learning based method to predict routing congestion in FPGA high-level synthesis and locate the highly congested regions in the source code. Based on the accurate congestion detection in HLS, routing congestion can be resolved easily at the early design stage. To the best of our knowledge, we are the first to build a congestion prediction model in FPGA high-level synthesis to help designers solve the routing congestion issues and optimize their applications. Our major contributions include: (1) To construct the dataset for model training, we develop an effective method to back trace the congestion metrics of CLBs and link with the HLS IR; (2) We propose seven informative categories of features and train multiple machine learning models to investigate the impact of our features. (3) Based on our trained model, the highly congested parts of the source code can be detected and we introduce several methods to reduce congestion in HLS and enhance design performance.
\vspace{-0.1cm}
\section{Motivation}\label{motivation section}
\begin{figure}
\vspace{-0.1cm}
  \centering
  \includegraphics[width=0.94\columnwidth]{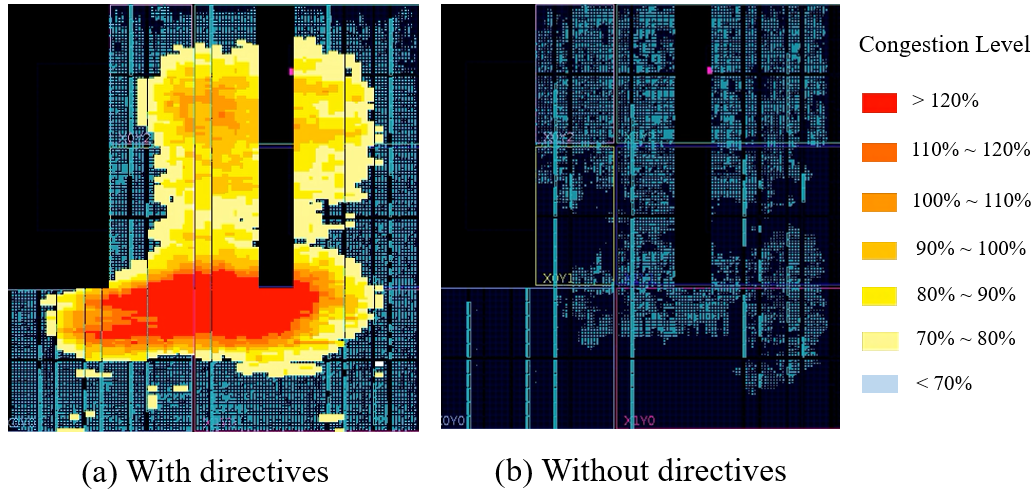}
  \vspace{-0.1cm}
  \caption{Congestion maps of \textit{Face Detection} from Vivado}
  \label{motivation}
  \vspace{-0.5cm}
\end{figure}

\vspace{-0.1cm}
We take an FPGA-based design \textit{Face Detection} from Rosetta \cite{zhou2018rosetta} as example to show the routing congestion issue in HLS. To meet the constraints of real-time applications, the designers apply several HLS directives such as function inlining, loop pipelining and unrolling, and array partitioning in \cite{zhou2018rosetta}. We compare two implementations on a Xilinx Zynq device (xc7z020clg484): one is the original implementation applied with the directives mentioned above, and the other one is the implementation without any directive applied. Figure \ref{motivation} shows the congestion maps from Xilinx Vivado and the color represents the degree of congestion. Congestion level denotes the percentage of routing resources used in corresponding tiles. A value over 100\% means that over 100\% of routing resources will be used and the router has to divert routes around that area. Table \ref{motivation_comp} compares multiple design performance metrics of the two implementations. The worst negative slack (WNS) indicates how much the design is missing the timing constraints.\\
\begin{table}
\vspace{-0.5cm}
\centering
\caption{\textsc{Performance Comparison}}
\vspace{-0.2cm}
\label{motivation_comp}
\begin{tabular}{|c|c|c|c|c|c|}
\hline
    {\tiny\textbf{Implementation}} &\tiny\textbf{\makecell{WNS(ns)}} &\tiny\textbf{\makecell{Max Freq.(MHz)}} & \tiny\textbf{\makecell{Latency(cycles)}} & \tiny\textbf{\makecell{Max Congestion(\%)}}\\
    \hline
    {\tiny{With Directives}} & \tiny{-13.643} & \tiny{42.3} & \tiny{1.08e+6} & \tiny{178.96} \\
    \hline
    {\tiny{Without Directives}} & \tiny{-0.066} & \tiny{99.3} & \tiny{1.73e+7} & \tiny{58.51} \\
    \hline    
\end{tabular}
\vspace{-0.2cm}
\end{table}
\indent We can see that implementations of the same design can vary greatly in performance if different optimization methods are adopted in HLS tools. By applying directives, the latency is reduced significantly, but the implementation becomes much more congested and the maximum frequency is decreased. For hardware designs, it is crucial to consider the trade-off among design performance metrics and select the right implementation method to meet the design constraints. However, current HLS tools like Vivado HLS fail to provide accurate estimations of physical metrics, since it is non-trivial to model the accumulated effects of each step in the RTL implementation flow. Although HLS can estimate the accurate clock cycle number, it is difficult to estimate the real timing (ns) and find the direction for further performance optimization of the design. At the same time, it takes too much time to repetitively run the complete C-to-FPGA flow to obtain the real timing in every round of design optimization. For example, it takes nearly seven hours to finish the logic synthesis and PAR for the \textit{Face Detection} application, compared to the significantly less time in HLS flow (several minutes). Therefore, it is of great benefit to provide early accurate prediction of physical information in HLS, such as routing congestion, which can guide the optimization and shorten the design cycle.\\
\indent For this reason, we propose our machine-learning based method to predict routing congestion during HLS and locate the highly congested regions in the source code. By identifying the problems and resolving congestion earlier, the maximum frequency is increased when then latency is still reduced, achieving a better trade-off to maximize the design performance using HLS tools. Note that the target of prediction in our problem is the routing congestion as shown in Fig. \ref{motivation}. 
\section{Congestion Prediction Flow}
\vspace{-0.1cm}
\begin{figure}
\vspace{-0.5cm}
  \centering
  \includegraphics[width=0.82\columnwidth]{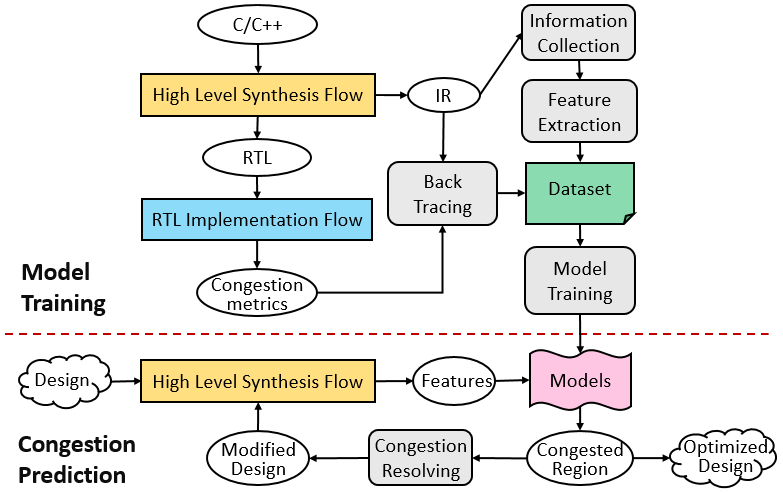}
  \caption{Overview of the proposed approach.}
  \label{overview}
  \vspace{-0.4cm}
\end{figure}
Figure \ref{overview} shows the overview of our proposed approach. There are two phases: one is to train the routing congestion prediction model and the other one is to predict the highly congested regions in the source code. During the training phase, with several HLS-based applications, we first run one time of the complete C-to-FPGA flow to obtain the routing congestion metrics. The C/C++ specifications of designs are synthesized into RTL models through the HLS flow, and then the RTL descriptions run through the implementation flow to generate the congestion metrics. Our approach begins with the intermediate representation (IR), which is generated by the HLS front-end, and targets at estimating congestion after the RTL implementation. The front-end compiler performs code optimization such as bitwidth reduction, which directly influences the data flow of generated RTL models\cite{coussy2009introduction}. By studying at the IR level, source-code transformations are considered and more accurate information can be extracted. To build the dataset for training, we back trace the routing congestion metrics per CLB to the operations in IR and collect necessary information from HLS to extract features. After that, features of each operation are extracted and the dataset is built. Each sample in the dataset consists of features of each operation and labels which are corresponding congestion metrics. With the dataset, we train machine learning models for congestion estimation. If the models are trained with a large dataset built from a variety of applications, routing congestion can be estimated accurately. If there are not many available applications to build a comprehensive dataset, the target design should go through the complete C-to-FPGA flow for one time to generate congestion metrics which will be used to enrich the dataset and improve the estimation accuracy. With the trained model, the highly congested regions in the source code of the target design can be detected during the prediction phase and users can resolve congestion issues in the HLS flow without running the time-consuming RTL implementation flow. 
\vspace{-0.1cm}
\subsection{Back Tracing and Information Collection}
\vspace{-0.1cm}
To obtain the dataset, we need to establish the one-to-one relationship between IR operations and congestion metrics and collect required information from the HLS synthesis flow. 
\begin{figure}
\vspace{-0.5cm}
  \centering
  \includegraphics[width=0.95\columnwidth]{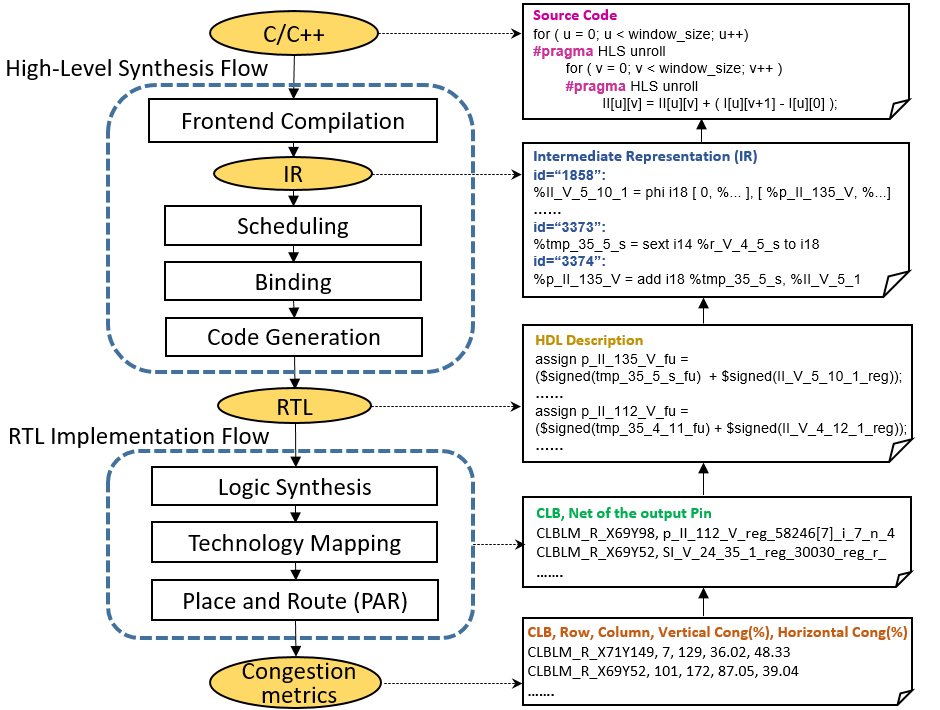}
  \vspace{-0.1cm}
  \caption{Back-tracing process.}
  \label{back-trace}
  \vspace{-0.7cm}
\end{figure}
\subsubsection{Back-tracing}
HLS synthesizes a design by scheduling IR operations to different control states and binding operations to functional units based on characterized libraries, as shown in Fig. \ref{back-trace}. It generates the RTL data path and FSM. After that, the RTL implementation flow synthesizes the HDL descriptions into gate-level netlists. The exact location of each operator on FPGA is unknown until PAR. Therefore, automatic back-tracing is employed to find the location of each IR operation on FPGA after PAR and extract corresponding congestion metrics of the CLBs in which the operation is implemented. Based on Xilinx Vivado design suite, our automatic back-tracing flow is shown in Fig. \ref{back-trace}, accompanied with code snippets at different steps. Physical information is extracted first, including vertical and horizontal congestion metrics per CLB and coordinates of row and column. For each CLB, the net names of the output pins of each cell are gathered through Vivado, which are then back traced to HDL descriptions. Finally, based on HLS-generated information during the synthesis procedure, we can establish the relationship between RTL operations and IR operations. Note that IR operations can be further back traced to the statements of the source code. 
\subsubsection{Information Collection}\label{info_collect}
\begin{figure}
\vspace{-0.4cm}
  \begin{minipage}[c]{0.16\textwidth}
      \caption{Merging the nodes that share the same RTL module.
    } \label{dependency}
  \end{minipage}\hfill
  \begin{minipage}[c]{0.3\textwidth}
   \vspace{-0.4cm}
   \includegraphics[width=\textwidth]{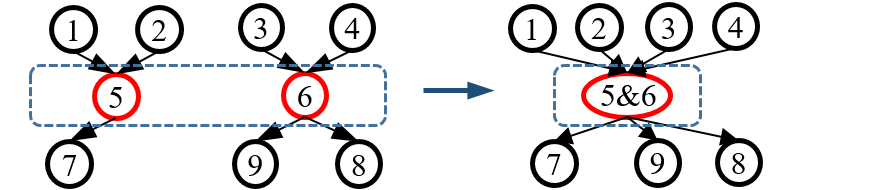}
  \end{minipage}
  \vspace{-0.8cm}
\end{figure}

To extract features that reflect the quality of hardware implementations like interconnection and resource usage, comprehensive HLS-based information is required to be collected in advance.\\
\indent \textbf{Information of each operator.} The values of multiple metrics for each operator are obtained from the HLS pre-characterization libraries. We collect the resource usage, operation type, bitwidth and delay (ns) for each operator.\\ 
\indent \textbf{Dependency among operators.} To trace the relationship between IR operations for one design, a dependency graph is constructed by storing each operation as one node and connecting dependent operations. The edge weight is measured by counting the number of wires for each connection. Take a 32-bit wide operator as an example, if one of its successors takes eight of the total 32 bits as the input signals, the actual number of wires for this connection is eight which is stored as the edge weight. Resource sharing is considered by replacing the operation nodes which share the same RTL module with one combined node, as shown in Fig. \ref{dependency}. The original nodes are removed and corresponding edges are redirected to the combined node. We also consider the influence of the function interface, which does not correspond to an IR operation but reflects the connections of I/O ports. Therefore, ``port" type nodes are also added to the graph to indicate which operators are connected to the same I/O port. \\
\indent \textbf{Scheduling and Global information.} HLS schedules operations into control states which reflect the overall latency of a design and the latest arrival time of the input signals for each operation. After keeping a record of the control states during which one operation is executed, exact operation latencies are obtained and the maximum distance between dependent operations could be measured. Detailed explanation is given in Section \ref{feature}. Global information is extracted based on HLS estimations, such as the resource usage and latencies of the functions and the total number of multiplexers.
\vspace{-0.2cm}
\subsection{Feature Extraction}\label{feature}
\vspace{-0.1cm}
To capture the characteristics of each operation in different designs, we extract 302 related features and divide them into seven categories, as listed in Table \ref{feature_list}. For better understanding, several categories are explained in detail below:
\subsubsection{Interconnection} The connections between operations reflect the local circuit complexity. As introduced in Section \ref{info_collect}, the edge weight in our dependency graph denotes the number of wires that connect the operators. For each operation, we compute the number of edges and sum the edge weight to represent the interconnection information, such as the number of predecessors and fan-in. Among all the connections of one operator, their contributions to fan-in/fan-out may vary, so we extract the connection with the maximum number of wires and compute its edge weight percentage of the total fan-in/fan-out. To provide a more comprehensive representation of the local relationships among operations, we further expand the scope and extract similar features after including the two-hop neighbors of each node in the graph. 
\subsubsection{Resource} Different kinds of resources on FPGA, i.e., DSP, BRAM, LUT and FF, occupy different locations of the device and hence constrain the placement of operations. Therefore, the resource-related information of operations and their surrounding operations for each resource type need to be considered as shown in Table \ref{feature_list}. 
Resource usage denotes the number of resource units consumed by each operation. Resource utilization ratios are computed by comparing the resource usage of one operation with the available resources on FPGA and the resource usage of the function in which this operation belongs to. Similar to the interconnection category, we extract resource-related features for each operation and compute the total resource usage and utilization ratios of the node's predecessors/successors. The two-hop neighbors of each operator are also considered.
\subsubsection{$\frac{\textit{\#Resource}}{\Delta\textit{T}_\textit{cs}}$} Besides resource-related metrics, the impact of surrounding operators is also related to the spatial distance. Take an operation $\textit{Op}$ with two different successors $\textit{S}_\textit{1}$ and $\textit{S}_\textit{2}$ as an example, if $\textit{S}_\textit{1}$ has to process the result of $\textit{Op}$ immediately after the computation finished, while $\textit{S}_\textit{2}$ is executed several clock cycles later, the distance constraints between $\textit{Op}$ and its two successors are different given the same target clock period. $\textit{S}_\textit{2}$ is allowed to be connected to $\textit{Op}$ through longer wires. Even if $\textit{S}_\textit{1}$ and $\textit{S}_\textit{2}$ consume the same amount of resources, they have different influence on the local layout around $\textit{Op}$. Therefore, we calculate $\frac{\textit{\#Resource}}{\Delta\textit{T}_\textit{cs}}$ to evaluate the combined effects of resource usage/utilization ratios and timing information. Given a pair of dependent operations and suppose the preceding operator outputs the computation results at control state $\textit{S}_\textit{p}$ and the subsequent operator begins to process the data at control state $\textit{S}_\textit{s}$, $\Delta\textit{T}_\textit{cs}$ is the subtraction of $\textit{S}_\textit{p}$ and $\textit{S}_\textit{s}$ based on the scheduling information as introduced in Section \ref{info_collect}. Similarly, we also compute this feature for the two-hop neighbors of each operation.
\begin{table}
\vspace{-0.5cm}
\renewcommand{\arraystretch}{0.9}
\centering
\caption{\textsc{List of Features}}
\vspace{-0.2cm}
\label{feature_list}
\begin{tabular}{|c|l|}
\hline
    \scriptsize\textbf{\makecell{Category}} &\scriptsize\textbf{\makecell{Feature Descriptions}} \\
    \hline
     \scriptsize{Bitwidth} & \scriptsize{Bitwidth of each operation.} \\
    \hline
     \multirow{5}{*}{\scriptsize{\makecell{Inter-\\connection}}} & \scriptsize{Fan-in and fan-out of each operator and their summation;} \\
    & \scriptsize{\#predecessors, \#successors and the summation;} \\
    & \scriptsize{The max. number of wires among all the connections to one-hop} \\
    & \scriptsize{neighbors and its percentage of the total fan-in and fan-out;} \\  
    & \scriptsize{Corresponding features after including two-hop neighbors.} \\
    \hline
     \multirow{6}{*}{\scriptsize{\makecell{Resource\\(for each \\resource\\type)}}} & \scriptsize{Resource usage and utilization ratios of each operation;} \\
    & \scriptsize{The total resource usage and utilization ratios of all the} \\
    & \scriptsize{predecessors and successors, and their summation;}\\
    & \scriptsize{The max. resource usage and the corresponding percentage }\\
    & \scriptsize{among all the one-hop neighbors;}\\
    & \scriptsize{Corresponding features after including two-hop neighbors.}\\
    \hline
      \scriptsize{Timing} & \scriptsize{Delay(ns) and latency (clock cycles) of each operation.} \\
    \hline
     \multirow{3}{*}{\footnotesize{$\frac{\textit{\#Resource}}{\Delta\textit{T}_\textit{cs}}$}} & \scriptsize{Resource usage and utilization ratios of predecessors/successors,} \\
    & \scriptsize{divided by the subtraction of control states $\tiny{\Delta\textit{T}_\textit{cs}}$;}  \\
    & \scriptsize{Corresponding features for two-hop neighbors.}\\
    \hline
     \multirow{2}{*}{\scriptsize{\makecell{Operator\\Type}}} & \scriptsize{The operation type of each operator, e.g., add, mul, xor, select;} \\
    & \scriptsize{The number of each kind of operations among one-hop neighbors.}\\
    \hline
     \multirow{6}{*}{\scriptsize{\makecell{Global\\Information}}} & \scriptsize{Resource usage of the top-level function ($\textit{F}_\textit{top}$);} \\
    & \scriptsize{Resource usage of the function in which the operation is located}\\
    & \scriptsize{(${\textit{F}_\textit{op}}$) and the corresponding percentage of the resources of ${\textit{F}_\textit{top}}$;}\\
    & \scriptsize{Target/estimated clock period and clock uncertainty of $\textit{F}_\textit{top}$ and $\textit{F}_\textit{op}$};\\
    & \scriptsize{Memories: \#words, \#banks, \#bits and \#primitives(words*bits*banks);}\\
    & \scriptsize{Multiplexers: number, resource usage, input size and bitwidth.}\\
    \hline
\end{tabular}
\vspace{-0.5cm}
\end{table}
\subsubsection{Global information} Global features reflect the possibility of congestion, such as the available resources on FPGA and resource usage of functions. We extract the resource and timing related metrics of the top-level function and the function in which the operation is located. We also extract the information of memories and multiplexers used in the functions because both of them are important resource components that influence the layout. Note that the total number of available resources on FPGA are not listed as features in Table \ref{feature_list}, because we focus on the same FPGA device currently. We consider the impact of the available FPGA resources when computing the utilization ratios in the resource category in Table \ref{feature_list}.
\vspace{-0.2cm}
\subsection{Model Training}
\vspace{-0.1cm}
\subsubsection{Sample Filtering}\label{filering section}
\begin{figure}
\vspace{-0.6cm}
  \begin{minipage}[c]{0.29\textwidth}
  \includegraphics[width=\textwidth]{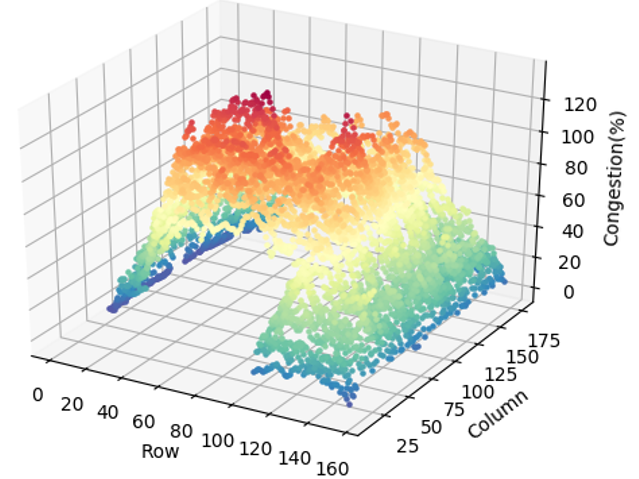}
  \end{minipage}\hfill
  \begin{minipage}[c]{0.17\textwidth}     
  \caption{Distribution of the vertical routing congestion metrics for \textit{Face Detection} on FPGA.} \label{filter}
  \end{minipage}
  \vspace{-0.5cm}
\end{figure}

When a loop is unrolled, multiple copies of the same operation will be generated and mapped to different hardware units, which are placed to different locations on FPGA after PAR. If the unrolling factor is large, the location of replicas can be distant. For example, in \textit{face detection} application\cite{zhou2018rosetta}, an unrolled loop generates 625 copies of the same set of operations, which are spread over the device. We find that parts of the replicas have similar features but their labels vary a lot because of their different physical locations. Figure \ref{filter} presents the distribution of vertical congestion metrics on FPGA for \textit{face detection} and shows that lower congestion metrics are distributed at the margin of the device compared to the higher values in the middle of FPGA. When some replicas are placed around the margin of FPGA, the corresponding samples may contain very small labels, which deviate from most of the replicas and influence the estimation accuracy. These operations, called as marginal operations, taking up around 3.4 percent of all the operations in our benchmarks, should be filtered and removed. After filtering the outliers, the trained model can predict routing congestion more accurately. Another reason for filtering marginal operations in an unrolled loop is that our target is to detect the most congested region in the source code and the marginal operations with very small congestion metrics do not influence the detection. 
\subsubsection{Machine-Learning Models} The selection of machine learning models largely depends on the problem to be solved and the data that is used. Therefore, we train and compare various regression models to show their differences in prediction of routing congestion on FPGA as is evident in Table \ref{accuracy}.\\
\indent \textbf{Linear Regression} is used to model a target which is expected to be a linear combination of input features. In this work, we apply the \textit{Lasso} linear model \cite{seber2012linear} with \text{L1}-regularization, which is to minimize the least-square penalty on the training data. The tuning parameter of the \textit{Lasso} model is a constant parameter that multiplies the L1-regularization term and determines the sparsity of model weights.\\
\indent \textbf{Artificial Neural Network (ANN)} is a promising technique to model non-linear relationships in the dataset\cite{yegnanarayana2009artificial}. 
Between the input and output layers, there are several hidden layers in which each neuron performs a weighted linear transformation on the values from the previous layer, followed by a non-linear activation function. It is more challenging to train the ANN model because a number of hyperparameters need to be tuned carefully and thus it takes a longer time to train.\\
\indent \textbf{Gradient Boosted Regression Trees (GBRT)} combines multiple weak prediction models to form a powerful regression ensemble, which is an accurate and effective learning method\cite{elith2008working}. GBRT builds the model in a stage-wise manner and introduces a weak estimator in each stage based on the gradients of the existing weak estimators. Several parameters require to be tuned such as the number of estimators and the learning rate that scales the step length of the gradient descent.
\vspace{-0.55cm}
\subsection{Resolving Congestion}
\vspace{-0.15cm}
Based on our trained models, routing congestion can be predicted and the most congested part of the source code can be recognized. Typically, it is easier to resolve congestion problems at the early design cycle. Through early detection, congestion can be resolved in advance without going through the downstream RTL implementation flow.
There are several methods to resolve routing congestion in HLS, such as modifying the code structure of the design and selecting suitable HLS directives. In Section \ref{case study}, we will present a detailed example and demonstrate how to resolve routing congestion problems easily at the source code level.
\vspace{-0.1cm}
\section{Experimental Results}
\vspace{-0.1cm}
\begin{table}
\vspace{-0.5cm}
\centering
\caption{\textsc{Property Summary of Benchmarks}}
\vspace{-0.2cm}
\label{design summary}
\begin{tabular}{|c|c|c|c|c|c|c|}
\hline
     \tiny{\textbf{Metrics}}&\tiny\textbf{\makecell{WNS(ns)}}
     &\tiny\textbf{\makecell{Freq.(MHz)}} &\tiny\textbf{\makecell{Vertical Cong(\%)}} & \tiny\textbf{\makecell{Horizontal Cong(\%)}} & \tiny\textbf{\makecell{Avg. (V, H) (\%)}}\\
    \hline
    {\textbf{\tiny{Max}}} & \tiny{-3.253} & \tiny{75.5} & \tiny{133.33} & \tiny{178.96} & \tiny{144.87}\\
    \hline
    \textbf{{\tiny{Min}}} & \tiny{-13.643} & \tiny{42.3} & \tiny{5.06} & \tiny{8.90} & \tiny{6.73} \\
    \hline
    \textbf{{\tiny{Avg.}}} & \tiny{-8.386} &\tiny{54.4} & \tiny{60.58} & \tiny{72.47} & \tiny{64.89} \\    
    \hline    
\end{tabular}
\vspace{-0.5cm}
\end{table}
We implement our approach in Python and train the models with the scikit-learn\cite{pedregosa2011scikit} machine learning library. Our dataset is built from the Rosetta\cite{zhou2018rosetta} benchmark suite, which contains six fully-developed realistic and optimized applications for HLS-based FPGA design. To fully utilize the available resources on FPGA and investigate the influence of routing congestion, we combine several benchmarks within the same top function and run the complete HLS and RTL implementation design flow to obtain the congestion metrics after PAR. Specifically, \textit{Face Detection} is complex enough to occupy the device and thus is tested individually. \textit{Digit Recognition} and \textit{Spam Filtering} are invoked by the same function and the rest three applications, namely, \textit{BNN}, \textit{3D Rendering} and \textit{Optical Flow}, are tested in an integrated function. We back trace the vertical and horizontal congestion metrics per CLB to the IR operations of each design, extract related features for each operation and build our dataset which consists of 8111 samples totally. Table \ref{design summary} summarizes the RTL implementation results of the benchmarks, including performance and routing congestion metrics. The average of vertical and horizontal congestion metrics in the last column is the mean value of the two metrics for each CLB. Routing congestion metrics denote the estimated utilization percentage of routing resources in the vertical and horizontal directions of the tiles on FPGA and a utilization rate of over 100\% indicates the potential routing problem around that region. All the designs are synthesized and implemented with Xilinx Vivado Design Suite 2018.1, and the target FPGA device is Zynq XC7Z020CLG484. The target frequency is set to 100 MHz. All experiments are conducted on an Intel Xeon CPU running at 2.5 GHz.
\subsection{Estimation Accuracy}
\vspace{-0.1cm}
We randomly select 80\% samples from our dataset for training and the rest 20\% for testing. We employ a 10-fold cross-validation on the training set and grid search is applied to find the best hyperparameters of each model. The testing set is totally unseen and only used to evaluate estimation accuracy. To evaluate the congestion estimation accuracy, we compute the mean absolute error (MAE), defined as $\frac{1}{N}{\sum|y_i-\hat{y}_i|}$, and the median absolute error (MedAE), defined as $\textit{median}(|y_1-\hat{y}_1|,...,|y_n-\hat{y}_n|)$. $\hat{y}_1$, $\hat{y}_i$ and $\hat{y}_n$ are the predicted values of the first, the $i$-th and the last samples and $y_1$, $y_i$ and $y_n$ are the corresponding actual values. $N$ is the size of the testing set. MAE measures the average value of the absolute relative errors and MedAE reflects the distribution of the absolute relative errors which is robust to outliers \cite{pedregosa2011scikit}.\\
\begin{table}
\vspace{-0.5cm}
\centering
\caption{\textsc{Congestion Estimation Results}}
\vspace{-0.25cm}
\label{accuracy}
\begin{tabular}{|c|c|c|c|c|c|c|c|}
\hline
    \multicolumn{2}{|c|}{\multirow{2}{*}{\tiny{\textbf{\makecell{Regression\\Models}}}}} &\multicolumn{2}{c|}{\tiny\textbf{\makecell{Vertical Cong (\%)}}} & \multicolumn{2}{c|}{\tiny\textbf{\makecell{Horizontal Cong (\%)}}} 
    & \multicolumn{2}{c|}{\tiny\textbf{\makecell{Avg. (V, H) Cong (\%)}}}\\
    \cline{3-8}
     \multicolumn{2}{|c|}{} & \tiny{MAE} & \tiny{MedAE} & \tiny{MAE} & \tiny{MedAE} & \tiny{MAE} & \tiny{MedAE}\\
    \cline{1-8}
    \multirow{3}*{\tiny{\makecell{Not\\Filtering}}} & {\tiny{Linear}} & \tiny{13.90} & \tiny{10.88} &\tiny{18.02} & \tiny{12.63} & \tiny{13.73} & \tiny{9.94}\\    
    \cline{2-8}
    &{\tiny{ANN}} & \tiny{12.19} & \tiny{7.91} & \tiny{17.68} & \tiny{12.62} & \tiny{12.27} & \tiny{8.17}\\
    \cline{2-8}
    &{\tiny{GBRT}} & \tiny{10.55} &\tiny{7.37} & \tiny{15.71} & \tiny{10.89} & \tiny{10.57} & \tiny{6.78}\\      
    \hline
    \multirow{3}*{\tiny{Filtering}} & {\tiny{Linear}} & \tiny{12.41} & \tiny{9.20} &\tiny{17.48} & \tiny{12.16} & \tiny{12.76} & \tiny{9.50} \\    
    \cline{2-8}
    &{\tiny{ANN}} & \tiny{10.23} & \tiny{7.43} & \tiny{16.61} & \tiny{11.78} & \tiny{11.67} & \tiny{7.83}\\
    \cline{2-8}
    &{\tiny{GBRT}} & \tiny{9.59} &\tiny{6.71} & \tiny{14.54} & \tiny{10.05} & \tiny{9.70} & \tiny{6.81}\\      
    \hline   
\end{tabular}
\vspace{-0.6cm}
\end{table}
\indent Table \ref{accuracy} compares the estimation accuracy of different models for routing congestion prediction. The GBRT model outperforms the ANN and the linear regression models and predicts all the three congestion metrics with the highest accuracy. As discussed in Section \ref{filering section}, we filter the marginal operations with the large deviation from other samples in an unrolled loop. As shown in Table \ref{accuracy}, filtering further improves the estimation accuracy for each model and reduces the MAE of the GBRT model to 9.59\%, 14.54\% and 9.70\% for the estimation of vertical, horizontal and their average routing congestion metrics, respectively. The smaller values of MedAE indicate that our models especially the GBRT model can predict routing congestion more accurately for at least half of the operations. Since we attempt to locate the most congested region in the source code, the accuracy of our model is sufficient to solve our problem and guide the subsequent optimization in HLS. 
In physical design, routing congestion is modeled to guide the placement with high accuracy\cite{chen2017ripplefpga,maaroufmachine}, because informative physical metrics can be obtained during the placement stage such as the number of bounding boxes, the net cuts per region and the pin count within a gcell. They require a highly accurate model to predict the routability, decide the location of each element and improve the quality of their placement algorithms. Compared to the works in physical design, the estimation error of our model is larger but acceptable due to the lack of detailed physical information and the wide gap of the abstraction level between HLS and the post-implementation after PAR. The cumulative optimization, imposed by each step in the downstream RTL implementation flow, is challenging to be captured at the source-code level. 
\vspace{-0.1cm}
\subsection{Feature Importance}
\vspace{-0.1cm}
To clarify and interpret different effects of our features on the routing congestion estimation, we assess the importance of different categories of features through the GBRT model. The GBRT model measures the feature importance by averaging the number of times that a feature is used as a split point of the trees in the ensemble model. Table \ref{feature importance} lists the most important categories of features for each congestion metric. The order of the feature categories in each column is determined by the ranking of importance. We can see that $\frac{\textit{\#Resource}}{\Delta\textit{T}_\textit{cs}}$ has the greatest influence on both vertical and horizontal congestion metrics and also affect the average congestion metric a lot. This is expected because $\frac{\textit{\#Resource}}{\Delta\textit{T}_\textit{cs}}$ reflects the combined impact of resource consumption and timing constraints and indicates the degree of crowding around operators. We also find that among the features of $\frac{\textit{\#Resource}}{\Delta\textit{T}_\textit{cs}}$ and resource categories, the features related to FF and LUT for the two-hop neighbors as explained in Section \ref{feature} exert greater influence. The interconnection category captures the complexity of connections between dependent operators, which directly impact the local layout. Global information indicates the overall performance of a design. Specifically, the related information of multiplexers and memories has a greater effect on congestion estimation than other global features. 
\begin{table}
\vspace{-0.5cm}
\renewcommand{\arraystretch}{0.4}
\centering
\caption{\textsc{Important Feature Categories}}
\vspace{-0.2cm}
\label{feature importance}
\begin{tabular}{|c|c|c|c|}
\hline
    \tiny{\textbf{Metrics}} &\tiny\textbf{\makecell{Vertical Congestion}} & \tiny\textbf{\makecell{Horizontal Congestion}} & \tiny\textbf{\makecell{Avg. (V, H) Congestion}}\\
    \hline
    \multirow{4}{*}{\tiny{\textbf{\makecell{Important\\ Feature\\Categories}}}} & \scriptsize{$\frac{\textit{\#Resource}}{\Delta\textit{T}_\textit{cs}}$} & \scriptsize{$\frac{\textit{\#Resource}}{\Delta\textit{T}_\textit{cs}}$} & \tiny{Resource}\\
     & \tiny{Resource} & \tiny{Resource} & \scriptsize{$\frac{\textit{\#Resource}}{\Delta\textit{T}_\textit{cs}}$} \\
     & \tiny{\makecell{Interconnection}} &\tiny{Interconnection} & \tiny{Interconnection} \\
     & \tiny{Global (Mux)} &\tiny{Global (Memory)} & \tiny{Global (Mux)} \\ 
    \hline    
\end{tabular}
\vspace{-0.2cm}
\end{table}
\subsection{Case Study}\label{case study}
\begin{figure}
\vspace{-0.6cm}
  \centering
  \includegraphics[width=0.9\columnwidth]{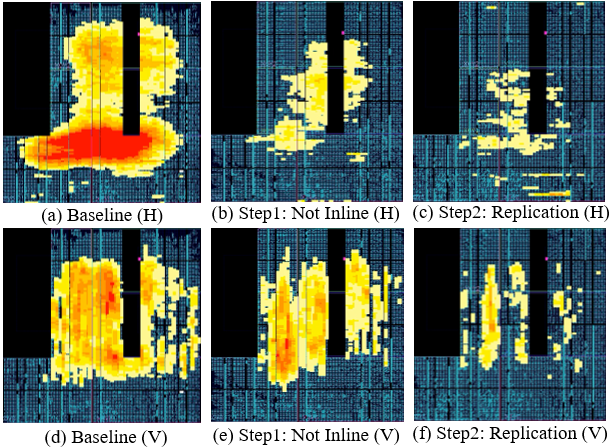}
  \vspace{-0.1cm}
  \caption{Resolving Routing Congestion of \textit{Face Detection}}
  \vspace{-0.5cm}
  \label{case solution}
\end{figure}

\vspace{-0.1cm}
When optimizing an HLS-based application, the trade-off relationship has to be considered carefully. As shown in Section \ref{motivation section}, \textit{Face Detection} suffers from severe routing congestion problem which degrades the maximum frequency significantly even though the latency is reduced. In this subsection, we illustrate how to improve the maximum frequency of \textit{Face Detection} and keep a low latency at the same time by resolving congestion. Based on our trained model, we can locate the most congested region and resolve congestion at the source-code level without running the RTL implementation flow.\\
\begin{table}
\vspace{-0.1cm}
\centering
\caption{\textsc{Case Study: Performance Improvement}}
\vspace{-0.2cm}
\label{case_comparison}
\begin{tabular}{|c|c|c|c|c|c|c|}
\hline
    {\tiny\textbf{Implementation}} &\tiny\textbf{\makecell{WNS \\(ns)}} &\tiny\textbf{\makecell{Max Freq.\\ (MHz)}} & \tiny\textbf{\makecell{$\Delta \textit{Latency}$\\(cycles)}} & \tiny\textbf{\makecell{Max Cong\\Vert, Hori (\%)}} & \tiny\textbf{\makecell{\#Congested CLBs\\$(\geqslant\textit{100\%})$}}\\
    \hline
    {\tiny{\textbf{Baseline}}} & \tiny{-13.643} & \tiny{42.3} & \tiny{1.08e+6} & \tiny{133.33, 178.96} & \tiny{1272}\\
    \hline
    {\tiny{\textbf{Not Inline}}} & \tiny{-3.504} & \tiny{74.1} & \tiny{+23} & \tiny{129.85, 97.60} &\tiny{193}\\
    \hline
    {\tiny{\textbf{Replication}}} & \tiny{-0.767} & \tiny{92.9} & \tiny{+0} & \tiny{106.15, 104.73} & \tiny{17}\\    
    \hline    
\end{tabular}
\vspace{-0.5cm}
\end{table}
\indent The results of performance improvement are shown in Table \ref{case_comparison}. The baseline is the original design applied with several optimization techniques, and ``Not Inline" and ``Replication" denote our methods applied in the first and second steps to reduce congestion respectively. The horizontal (H) and vertical (V) congestion maps in each step are presented in Fig. \ref{case solution} and the color represents the same meaning as in Fig. \ref{motivation}. In \textit{Face Detection}, each image unit goes through the cascade classifier function which contains multiple stages of classifiers. The cascade classifier function and all the classifiers within it are inlined to reduce the latency. However, this incurs the issue since \textit{function inlining} increases the complexity in C synthesis and generates a larger design. Routing congestion is detected at the region where multiple results returned by the classifiers are summed up and compared. Therefore, the first step is to remove \textit{function inlining} which significantly reduces the maximum congestion metrics and the number of congested CLBs, as shown in Table \ref{case_comparison}. The performance is also improved with a higher maximum frequency and a slightly increased latency. After the first step, large routing congestion metrics are detected at the inputs of classifiers. This is because all the classifiers access data from the same completely partitioned array and multiple classifiers share the same inputs, leading to a large number of interconnections. To reduce the number of interconnections, the second step is to modify the high level source code by replicating the values of the input data and sending the copies to different classifiers. After that, the number of congested CLBs with over $100\%$ congestion metrics is reduced to 17 and the maximum frequency is increased to 92.9MHz, while the latency remains unchanged compared to the step one. By modifying the source code or selecting suitable HLS directives, routing congestion is resolved easily and the performance is improved significantly, demonstrating that early detection of routing congestion in HLS is of great importance to enhance the developing efficiency.
\vspace{-0.1cm}
\section{Conclusion}
\vspace{-0.03cm}
In this paper, we propose a novel machine-learning based method to predict routing congestion for FPGA high-level synthesis. After back tracing congestion metrics to IR operations, we extract informative features and train three different machine learning models for congestion estimation. Experiments show that our GBRT model achieves the highest prediction accuracy. Based on the model, routing congestion can be reduced easily and performance is improved significantly.

\vspace{-0.1cm}
\bibliographystyle{IEEEtran}
\bibliography{sig}

\end{document}